# Enhanced Li capacity at high lithiation potentials in graphene oxide

Maria E. Stournara, Vivek B. Shenoy*

**Abstract**

We have studied lithiation of graphene oxide (GO) as a function of oxygen coverage using first principles calculations. Our results show that the lithiation potentials and capacities in GO can be tuned by controlling the oxygen coverage, or the degree of reduction. We find a range of coverages where the lithiation potentials are above the solid electrolyte interface (SEI) formation threshold, but with capacities comparable to, or larger than graphite. We observe that in highly oxidized and mildly reduced sheets, lithiation occurs through the formation of Li-O bonds, whereas at low coverages that are typical of reduced-GO (rGO) (O:C $\sim$ 12.5 %), both Li-O bonds and $LiC_6$ configurations are observed. The covalent Li-O bond is much stronger than the bonds formed in the $LiC_6$ ring and the lithiation potentials for epoxides at high and medium coverages are generally large ($> 1$ eV). For these congifurations, as in the case of $Li_4Ti_5O_{12}$ anodes, there will be no formation of SEI, but with the additional advantage of having higher lithium storage capacity than $Li_4Ti_5O_{12}$. In reduced GO sheets, the presence of residual oxygen atoms allows for formation of covalent Li-O bonds that lead to storage capacities and lithiation potentials higher than that of graphite. Finally, our calculations show high lithiation potentials for the edges of graphene nanoribbons, which will impede the formation of SEI and hence lead to large reversible capacity.

*Corresponding author: Vivek_Shenoy@brown.edu; 401-863-1475.

# 1. Introduction

As the global energy demand increases, developing energy storage systems with higher energy densities is becoming more and more critical. Rechargeable Li-ion batteries have been widely used in portable electronics due to their high gravimetric energy storage and are now being used in light vehicles. However, for heavy vehicles which demand delivery of much higher currents, the need for batteries with higher energy density and capacity is of great importance [1]. Furthermore, since batteries in electric vehicles (EVs) and hybrid EVs are expected to have high lifetimes, aging of Li-ion cells is becoming more and more of a concern. In terms of cycle life, a lifetime up to 3000 deep cycles is generally desirable for high-energy applications typical for EVs. These stringent requirements on enhanced gravimetric capacity of Li-ion batteries together with the need for prolonged lifetime, poses a great challenge from the perspective of fundamental science and materials technology.

Graphite is the most commonly used anode material in rechargeable Li-ion batteries due to its high Coulombic efficiency (the ratio of the extracted lithium to the inserted lithium). However, its gravimetric capacity is relatively low (theoretical value of 372 mAh/g), since every six C atoms can host only one Li ion by the formation of the intercalation compound $LiC_6$ with a lithiation potential of about 0.5 V vs. Li/Li+ [2, 3]. This potential is lower than the onset potential for the formation of the solid electrolyte interface (SEI) films (typically about 0.8V for a 1.2 M $LiPF_6$ in a mixture of ethylene carbonate/diethylene carbonate electrolyte) [4, 5]. Subsequently, Li from the anode has to first form the SEI layer (typically of the order of 10 nm thickness) before it intercalates into graphite particles [4, 5] in the anode. While SEI film is a necessity because of electrolyte instability at low potentials in systems like graphite, the damage that occurs to the SEI layer during subsequent cycling (and the resulting reformation) causes loss of cycleable lithium ions and hence capacity fade and shortening of the lifetime of the battery [6, 7, 8]. Clearly, the performance of state-of-art batteries can be improved by identification of graphite-based materials or derivatives that a) have higher capacity than graphite itself and b) whose lithiation potentials are above 0.8 V vs. Li/Li+ (so that the anodes can function without the need for the formation of SEI). The purpose of this study is to use ab initio density functional theory (DFT) calculations to identify graphene oxide (GO) structures to achieve these goals. Note that lithium titanate ($Li_4Ti_5O_{12}$) is an anode material with lithiation potential of about 1.5 V vs. Li/Li+, but with capacity lower than that of graphite (167 mAh/gm)[9, 10]. As we discuss below, GO systems exhibit potentials comparable to $Li_4Ti_5O_{12}$, but with significantly higher capacity.

Our work is motivated by recent experimental studies that suggest that the presence of Faradaic reactions at oxygen-containing functional groups in nanostructured carbon electrodes improves the gravimetric capacity of the anode [11] and the work of Lee et al., that report high-energy and power capabilities of layer-by-layer multi-wall carbon nanotube (LBL-MWNT) electrodes due to redox reactions involving oxygen groups bound to the edges [12]. While experiments suggest superior performance metrics of oxygen-functionalized graphene, to the best of our knowledge, there has not been any theoretical work on the lithiation of GO. Since GO is not stoichiometric and it is highly hygroscopic, its composition (oxygen content) can vary with the synthesis method and the ambient. The oxygen functional groups on the basal plane are generally thought to consist of epoxide and hydroxyl molecules, although evidence for the presence of ketones and phenols has been found in recent work [13], while edges generally comprise of carboxyls and ketone. The oxygen content in GO can be controlled by thermal treatment as well as by employing chemical agents such as hydrazine and alcohols. Hence, the following questions must be addressed to assess the viability of GO as an anode material: How do the lithium storage capacity and the lithiation potential of GO depend on the oxygen content? Is there any particular oxygen concentration that would lead to maximum capacity ? In this study, we address these questions using DFT ab initio simulations of lithiation of GO.

## 2. Methodology

All our calculations are performed using the Vienna Ab Initio Simulation Package (VASP) [14] with the Projector Augmented Wave (PAW) [15, 16] method and the Perdew-Burke-Ernzerhof (PBE) [17] form of the generalized gradient approximation (GGA) for exchange and correlation. From convergence studies, we determined a kinetic energy cut-off of 500 eV and a Fermi smearing width of 0.05 eV. For the two dimensional (2-D) Brillouin zone of GO, we used (15 × 30) and (10 × 15) Monkhorst- Pack grids and a 20 and 40 uniform k-point grid for the one dimensional (1-D) nanoribbons. The pe-riodic images of the zigzag graphene nanoribbons (ZGNRs) were separated by 12 Å of vacuum in the non-periodic directions. All atoms and cell-vectors are relaxed with a force tolerance of 0.02 eV.

To obtain the lithiation potential, V, consider the reaction Li +GO (anode) ⇌ LiGO (lithiated anode); V, as described in the past for a variety of electrode materials [18, 19] is:

$$V = -\frac{\Delta G_f}{z \cdot F}, \qquad (2.1)$$

where the change in Gibb's free energy is

$$\Delta G_f = \Delta E_f + P\Delta V_f - T\Delta S_f. \qquad (2.2)$$

Since the term $P\Delta V_f$ is of the order of $10^{-5}$ electron volts [18, 19], whereas the term $T\Delta S_f$ is of the order of the thermal energy (26 meV at room temperature), the entropy and pressure terms can be neglected and the free energy will be approximately equal to the formation energy $\Delta E_f$ obtained from DFT calculations. The formation energy is defined as

$$\Delta E_f = E_{Li_xGO} - (xE_{Li} + E_{GO}) \qquad (2.3)$$

where $x$ is the number of Li atoms inserted in the computational cell, $E_{Li_xGO}$ is the total energy of the $Li_xGO$ structure, $E_{Li}$ is the total energy of a single Li atom in elemental body-centered cubic Li, and $E_{GO}$ is the total energy of a particular GO structure. If the energies are expressed in electron volts, the potential of the $Li_xGO$ structures vs Li/Li+ as a function of lithium content can be obtained as [18, 19]

$$V = -\frac{\Delta E}{x} \qquad (2.4)$$

The composition range over which Li can be reversibly intercalated determines the battery capacity.

## 3. Results and discussion

Before we study the lithiation of GO as a function of O-coverage, we first consider the formation energy of the Li-O bonds on commonly found functional groups, namely epoxides and carbonyls. In Fig.1 we have considered lithiation of an isolated epoxide, a pair of epoxides in two different arrangements and a "hole" formed by a carbonyl pair. These calculations show that the lithiation potential for attachment to an oxygen atom is about 1.7 V for epoxide pairs, indicating a strong Li-O bond compared to the bonds in the $LiC_6$ ring. The small differences of the order of a fraction of an eV between the different cases can be attributed to the interactions between the oxygen groups which lead to the stability of certain arrangements of oxygen atoms. Here, for a pair of oxygen atoms on the unlithiated sheet, carbonyls are the most stable configuration [13], leading to a smaller lithiation potential compared to the epoxide pairs. The Li-O bond lengths vary from 1.716 $\mathring{A}$ for lithiated epoxide pairs to 1.74 $\mathring{A}$ for single epoxides.

Next, we consider the case of GO sheets with high oxygen coverages. Our calculations show that lithiation of graphene sheets with only epoxide groups with O:C ratios of 50 % and 66.6 %, leads to fracture of C-C bonds and disintegration of the structure. Hence, at high oxygen content we consider the oxygen groups to be present in the form of both epoxides and hydroxyls (Fig.2). Specifically, we study the cases, $C_{12}O_4(OH)_4$ (O:C = 66.6 %) and $C_8O_2(OH)_2$ (O:C = 50 %). When only the epoxides in these structures are lithiated (Fig.2a and b), due to the formation of the stability of the Li-O bonds, we find lithiation potentials to be higher than 0.8 V vs. Li/Li+. While the lithiation potential for these structures implies no formation of SEI, their capacity is either comparable (Fig.2a) to, or smaller (Fig.2b) than graphite, but signifantly higher than $Li_4Ti_5O_{12}$ [9, 10]. When the hydroxyl groups in $C_8O_2(OH)_2$ are lithiated, the capacity increases to 441 mAh/g. However, since Li-O bond is less stable than O-H bond, the potential reduces to 0.585 V. It is important to note that in this case we obtain somewhat higher lithiation potentials and capacities compared to graphite.

In the case of mildly reduced GO sheets with oxygen coverages in the range 25-37.5 % [20, 21], lithiation occurs exclusively by the formation of O-Li bonds as shown in Fig.3 and Fig.4. Here, the oxygen coverages are still large enough to allow room for formation of the $LiC_6$ rings, which is this the reason why the lithiation potential goes below 0.8 V vs. Li/Li+ (closer to that of $LiC_6$) at higher capacities. We consider only $C_{12}O_1(OH)_2$ and $C_8O_2$ for 25 % and $C_8O_2(OH)_2$ and $C_8O_3$ configurations for 37.5 % coverage. Although other structures are possible [22], our calculations show that these are the only structures that survive upon lithiation; large strains in the lithiated structures lead to fracture of C-C bonds in other structures. As in the case of highly oxidized sheets, lithiation of epoxide groups results in high lithiation potentials (> 1.4 V vs. Li/Li+ ) , whereas additional lithiation of hydroxyls improves the capacity while decreasing the potential to values below 0.8 V vs. Li/Li+ (Fig.3). When all the oxygen atoms are lithiated, the lithiation potential drops to 0.60 V vs. Li/Li+, but with a capacity that is ∼ 25 % more than that of graphite. Interestingly, as shown in Fig.4, at 25 % O coverage we find that there is one configuration totally based on epoxides, for which the system will operate without SEI formation (the lithiation potential is 1.59 V vs. Li/Li+) at the same capacity as graphite (377 mAh/g). For mixed epoxide-hydroxyl configurations the potential may increase up to 2.16 V vs. Li/Li+, but with a significant decrease of the capacity to values even lower than the capacity of $Li_4Ti_5O_{12}$. Less suitable anode configurations result from additional lithiation of hydroxyls, which lead to almost zero potential at low capacity.

It is well known that all the oxygen atoms on the graphene basal plane cannot be removed even with the harshest of the reduction methods [23]. Reduced-GO (rGO) typically contains residual oxygen atoms with a concentration of 12.5 % (O:C ratio), that are $sp^3$ bonded to approximately 20% of the carbon atoms [21, 24, 25]. To model lithiation of rGO, we consider structures with 8.3-11 atomic percentage of oxygen, meaning 12.5 % O:C. In Fig.5, we present the initial and lithiated configurations in the order of decreasing potential. Our simulations show that Li atoms can attach to rGO both by forming bonds with oxygen atoms and by forming $LiC_6$ rings as in the case of graphite. This combined lithiation mechanism allows rGO to exhibit both greater potentials and capacities compared to GO-Li structures with higher oxygen content. In particular, Fig.5 demonstrates that at potentials close to the threshold potential for SEI formation, we find two structures that have higher capacity than graphite. Futhermore, as shown in Fig.6, the capacity of rGO can be as high as two times the capacity of graphite at lower lithiation potentials that are comparable to that of graphite. Our results for the lithiation potentials at different O coverages are presented in Fig.7; for all the cases where epoxides are lithiated, we find lithiation potential above the SEI formation threshold, > 0.8 V vs. Li/Li+.

Finally, we consider lithiation of the ketone groups at the edges of ZGNRs (Fig.8) to model the experiments on charging of LBL-MWNT. Here, we have considered two different edge coverages of O atoms. In both cases, we find high lithiation potentials (> 0.8 V vs. Li/Li+) due to the strong bonds that Li atoms form with the edge-oxygen species. The absence of SEI in this case, suggested by our computed potentials, can potentially explain the very high first-cycle efficiency observed in the experiments [12].

Our studies here are confined to lithiation of single GO sheets. However, since GO is hydrophilic, interlamellar water molecules are always present in interlayer voids even after prolonged drying and the spacing between the individual sheets is generally in the range 8-12 $\mathring{A}$. Therefore, Li+ should be able to diffuse through the interlayer voids and attach to the individual sheets that we have considered here. We also note that inspite of the presence of voids, these materials display remarkable mechanical properties with tensile moduli in the range of 6-42 GPa and fracture strengths of 100-132 MPa [26]. In addition, in this paper we have confined attention to graphene oxide sheets with periodic arrangement of functional groups considered in previous theoretical work [27, 28]. While the arrangement of functional groups on the basal plane of GO is indeed random in realistic conditions, as we have shown, the lithiation potentials and capacity is determined by the relative fractions of the Li-O and LiC$_6$ bond that depends to a large extent on the overall coverage. Therefore, we do not expect that they will be significantly different for periodic and random arrangement of functional groups for a given coverage.

## 4. Conclusions

In summary, we have performed first principles calculations to study the trends of lithiation potentials and gravimetric capacities in the case of GO as a function of oxygen coverage and for ZGNRs decorated with oxygen atoms at the edges. Our simulations show that Li atoms can attach to GO by either forming bonds with oxygen atoms or by forming LiC$_6$ rings as in the case of graphene. We find that the former is prevalent at high and medium oxygen coverages (C:O 50-66.6 % and C:O 25-37.5 %) ,whereas both these configurations are observed at low coverages (12.5 % O:C), typical of rGO. Since Li-O bonds are much stronger than the bonds in the LiC$_6$ rings, the lithiation potentials at high O-coverages is generally large (> 1 V vs. Li/Li+). While the capacity at high coverages is smaller than that of graphite, it is bigger than the capacity of non-SEI forming Li$_4$Ti$_5$O$_{12}$ anodes. On the other hand at low O-coverages as in the case of rGO, where Li can attach to both to oxygen groups and in the form of $LiC_6$ rings, the capacity can be greater than those at high O-coverages, with lithiation potentials that are lower yet bigger than the potential for SEI formation. Our simulations show that there is a O-coverage window (close to 12.5 % O:C), where the best performance characteristics can be achieved. This result is consistent with the charge-discharge curves for GO reported in Ref. 11. For the case of ZGNRs the lithiation potentials are relatively high due to the strength of the C-O bonds, which will impede the formation of SEI and hence lead to large first-cycle capacity.

**Acknowledgments**

VBS thanks Dr. V. Sethuraman for a number of insightful discussions and suggestions. We gratefully acknowledge support from the NSF and the Brown University MRSEC program.

| GO structure | Lithiated structure | Lithiation potential vs. Li/Li+ (V) |
|---|---|---|
| 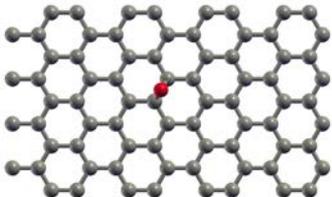 | 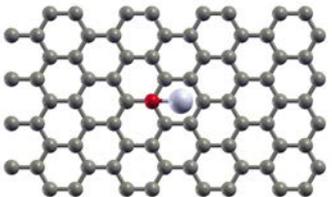 | 1.663 |
| 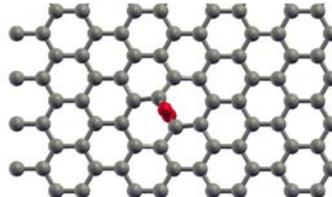 | 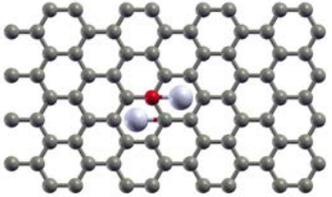 | 1.283 |
| 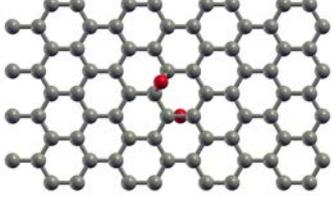 | 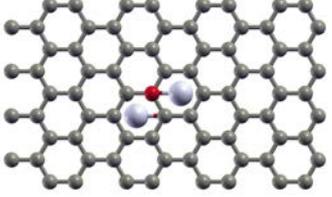 | 1.698 |
| 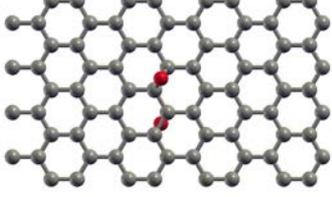 | 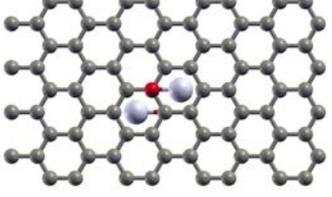 | 1.725 |

**Figure 1:** Atomic structures of oxygen atoms on the graphene basal plane and corresponding lithiated structures along with the intercalation potentials.

| GO structure | Lithiated structure | % O:C Atomic ratio | Lithiation potential vs. Li/Li+ (V) | Gravim. capacity ($\frac{mA\cdot h}{g}$) |
|---|---|---|---|---|
| 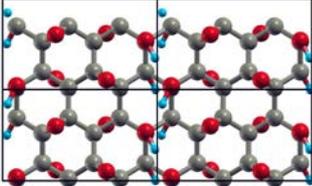 | 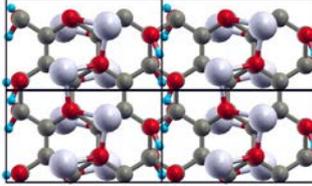 | 66.6 | 2.081 | 352 |
| 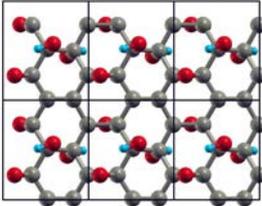 | 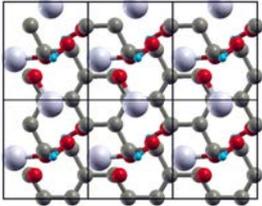 | 50 | 1.882 | 224 |
| 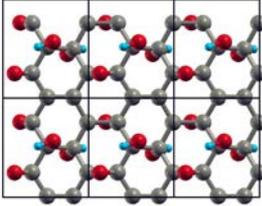 | 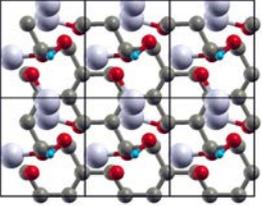 | 50 | 0.585 | 441 |

**Figure 2:** For 50-66.6% O:C GO sheets, lithiation of the epoxides results in potentials that are higher than 0.8 V vs. Li/Li+ and capacities that are comparable to, or smaller than graphite, but signifantly higher than $Li_4Ti_5O_{12}$. The unit cells considered in this and the following figure sets are represented by orthogonal boxes.

| GO structure | Lithiated structure | Lithiation potential vs. Li/Li+ (V) | Gravim. capacity ($\frac{mA\cdot h}{g}$) |
|---|---|---|---|
| 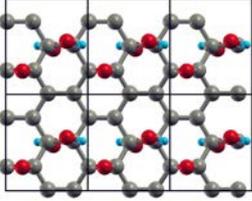 | 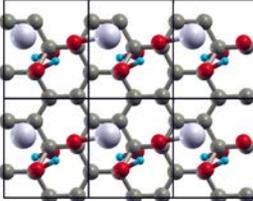 | 1.494 | 252 |
| 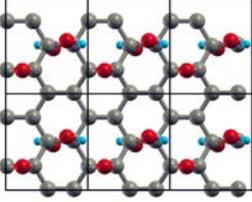 | 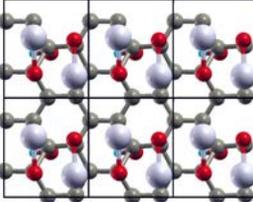 | 0.706 | 336 |
| 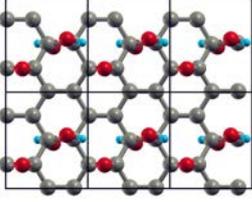 | 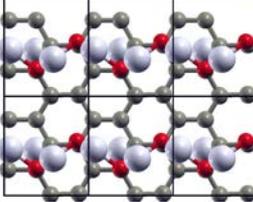 | 0.609 | 486 |

**Figure 3:** For 37.5% O:C GO sheets, as in the case of 50-66.6% O:C GO sheets, lithiation of epoxide groups results in high lithiation potentials (> 1.4 V vs. Li/Li+), whereas additional lithiation of hydroxyls improves the capacity while decreasing the potential to values below 0.8 V vs. Li/Li+.

| GO structure | Lithiated structure | Lithiation potential vs. Li/Li+ (V) | Gravim. capacity ($\frac{mA \cdot h}{g}$) |
|---|---|---|---|
| 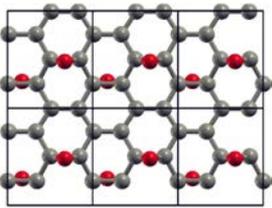 | 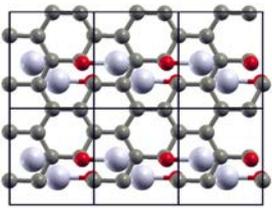 | 1.594 | 377 |
| 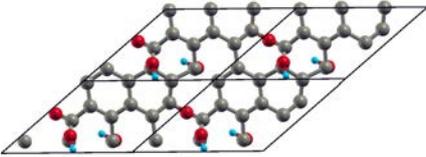 | 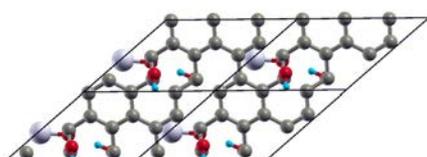 | 2.161 | 133 |
| 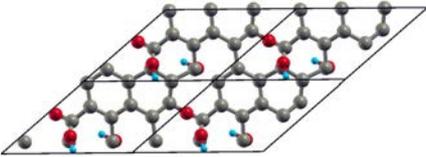 | 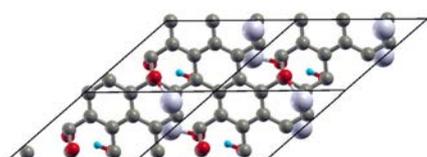 | 0.053 | 258 |

**Figure 4:** For 25 % GO lithiated sheets totally based on epoxides, the system will operate without SEI formation (the lithiation potential is 1.59 V vs. Li/Li+) at the same capacity as graphite (377 mAh/g).

| GO structure | Lithiated structure | Lithiation potential vs. Li/Li+ (V) | Gravim. capacity ($\frac{mA \cdot h}{g}$) |
|---|---|---|---|
| 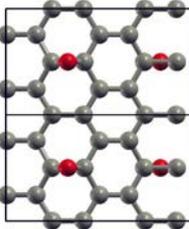 | 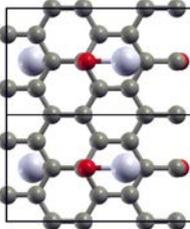 | 1.806 | 224 |
| 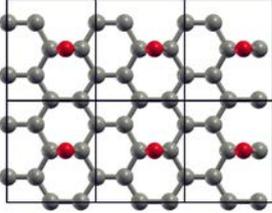 | 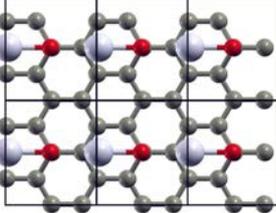 | 1.421 | 224 |
| 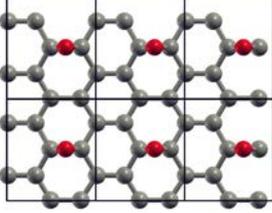 | 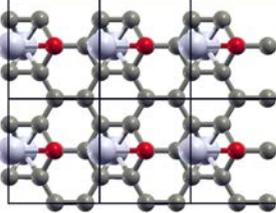 | 0.960 | 425 |
| 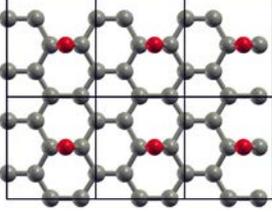 | 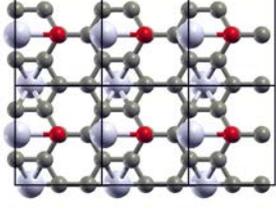 | 0.883 | 425 |
| 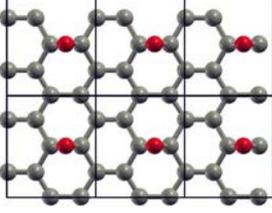 | 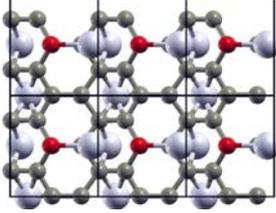 | 0.594 | 604 |

**Figure 5:** For 12.5 % O:C GO sheets, a combined lithiation mechanism allows rGO to exhibit both greater potentials and capacities compared to GO-Li structures with higher oxygen content.

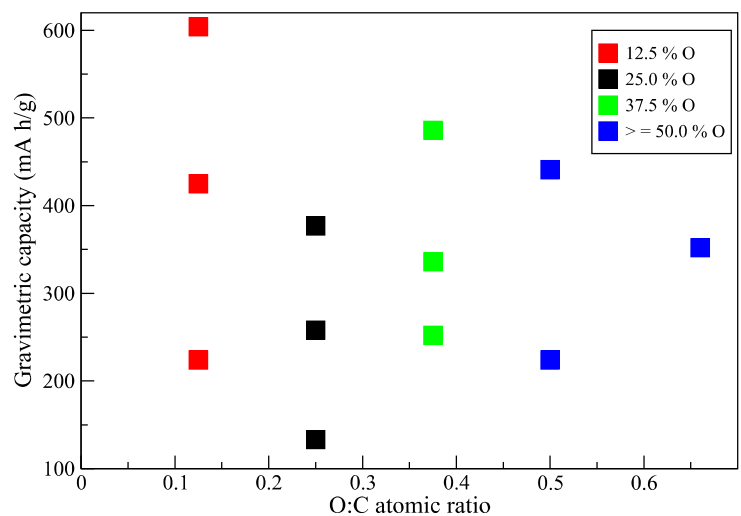

**Figure 6:** Gravimetric capacities for different O:C atomic ratios. The dashed line corresponds to the capacity of graphite anodes. The capacity of rGO can be as high as two times the capacity of graphite.

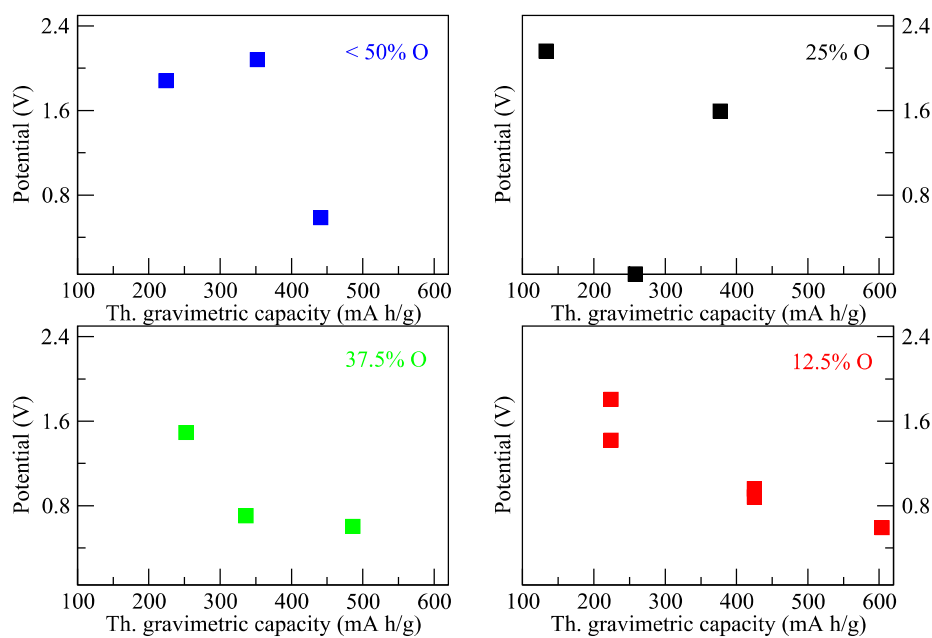

**Figure 7:** Potential versus capacity for lithiated structures at several O:C atomic ratios. For all of the cases where epoxides are lithiated, the lithiation potential is above the SEI formation threshold (horizontal dashed line at > 0.8 V vs. Li/Li+). At low O-coverages the capacity can be significantly greater than graphite (vertical dashed line at 372 mAh/g).

| GO structure | Lithiated structure | Lithiation potential vs. Li/Li+ (V) |
|---|---|---|
| 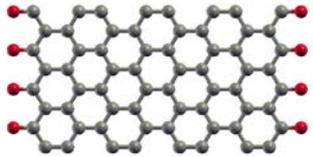 | 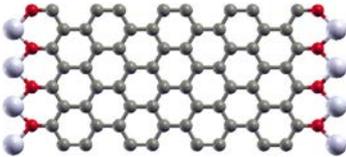 | 1.670 |
| 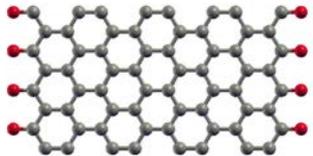 | 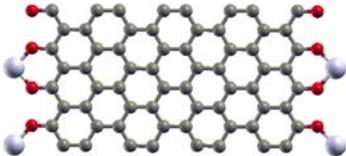 | 1.384 |

**Figure 8:** Lithiated ZGNRs exhibit high lithiation potentials (> 0.8 V vs. Li/Li+) due to the strong bonds that Li atoms form with the edge-oxygen species.